\documentclass[a4paper,11pt]{article}
\usepackage{jheppub} 
\usepackage{lineno}
\usepackage{graphicx,amsfonts,amsmath,amssymb,epsfig,color, mathrsfs,latexsym,url,wasysym,float,xfrac}
\usepackage{subcaption}

\newcommand{\be}{\begin{equation}}
\newcommand{\ee}{\end{equation}}

\newcommand{\ba}{\begin{eqnarray}}
\newcommand{\ea}{\end{eqnarray}}

\newcommand{\beq}{\begin{equation}}  \newcommand{\eeq}{\end{equation}}

\title{\boldmath On Measuring Distances in the Quantum Gravity Landscape}

\author{Amineh Mohseni$^{a,c}$,}
\author{Miguel Montero$^b$,}
\author{Cumrun Vafa$^a$,}
\author{Irene Valenzuela$^{b,c}$}

\affiliation[a]{Jefferson Physical Laboratory, Harvard University, Cambridge, MA 02138, USA}
\affiliation[b]{Instituto de F\'{i}sica Te\'{o}rica UAM-CSIC and Departamento de F\'{i}sica Te\'{o}rica, Universidad Aut\'{o}noma de Madrid, Cantoblanco, 28049 Madrid, Spain}
\affiliation[c]{CERN, Theoretical Physics Department, 1211 Meyrin, Switzerland}

\preprint{IFT-24-097\ , \ 
CERN-TH-2024-101}

\emailAdd{amohseni@g.harvard.edu, miguel.montero@csic.es, vafa@g.harvard.edu, irene.valenzuela@cern.ch}

\abstract{In this note, we propose a generalized notion of distance between vacua in the theory of a scalar field $\phi$ with scalar potential $V(\phi)$ coupled to gravity. 
We propose the  normalized tension of domain wall connecting different field values, with a varying normalization relative to a local energy scale, as the distance.  We show this definition reproduces the usual moduli space distance for zero potential, as well as the $d\propto |\log \Lambda|$ behavior with the vacuum energy $\Lambda$ in the AdS case, previously proposed in the literature.  In the case of large AdS  we also obtain the expected exponent of mass versus distance in one particular case, when the mass of the light tower is $m\sim \sqrt \Lambda$ and there is a single extra dimension decompactifying. We also discuss the features and shortcomings of alternative but related proposals.}

\begin{document}
    \maketitle
    \flushbottom
\section{Introduction}

Quantum gravity theories possess particular features, which are the object of study of the Swampland program \cite{palti2019swampland,van2022lectures,agmon2022lectures}.   An important subset of these pertains to the physics of asymptotic regimes in field space.  For example at large distances in the moduli space of Minkowski theories one expects to get an exponentially light tower of particles
\cite{ooguri2007geometry}, where distances are measured using the kinetic term of the scalars in the Einstein frame action as a metric. The same situation is expected to continue holding in the case with non-vanishing vacuum energy, although defining a proper notion of distance in that case is more challenging. If we consider two AdS vacua connected by a moduli space, we can still use the Zamolodchikov distance of the conformal manifold of the dual Conformal Field Theory to recover the exponential drop-off of the tower of states as distance becomes parametrically large  \cite{Baume:2020dqd,Perlmutter:2020buo,Baume:2023msm,Ooguri:2024ofs}.  The notion of distance has also been proposed in the AdS/dS case  \cite{lust2019ads} with a number of followup works \cite{Kehagias:2019akr,DeBiasio:2022nsd,DeBiasio:2022omq,Velazquez:2022eco,DeBiasio:2020xkv,DeBiasio:2022zuh,Shiu:2022oti,Shiu:2023bay,Li:2023gtt,Basile:2023rvm,Palti:2024voy} including its interplay with the asymptotic towers of states.
Most of these notions of distance in field space are motivated by the kinetic term in the action and do not take into account the existence of potential as a functions of fields (see, however, \cite{Basile:2023rvm} and \cite{Palti:2024voy} for an example that also considers the potential, and \cite{schimmrigk2018swampland} for an earlier attempt to account for the effect of the potential).  One of the main aims of this paper is to make progress in filling this gap.

Defining a notion of distance between different quantum field theories is certainly a long-standing open challenge, and interesting proposals have been investigated in the literature 
(see e.g.\cite{o1993geometry,dolan1998renormalisation, anselmi2011distance,douglas2013spaces, Bachas:2013nxa, Balasubramanian:2014bfa,Stout:2022phm}).
In this work, we will take inspiration from the Cobordism conjecture \cite{mcnamara2019cobordism} to define a distance between theories. This Swampland constraint implies the existence of a domain wall of finite tension connecting every two quantum gravity theories of the same dimension. This is proposed to avoid the existence of global symmetries, which cannot be exact in quantum gravity. Hence, can we use the domain wall tension (or some related quantity) to define a distance between the theories? If so, this would connect nicely with another quantum gravity expectation: the finiteness of the quantum gravity landscape. If all theories are connected by finite tension domain walls, they would all be at finite distance from each other, implying they are all part of a single, universal quantum theory of gravity.

Given two effective field theories (EFTs), we can consider the path of minimal euclidean action in the space of field configurations (for short: the configuration space) that interpolates between the two EFTs. Whenever this euclidean solution admits a ``thin wall' description, the action can be written in terms of the domain wall tension. We will use this euclidean action (normalized by the euclidean energy density of the configuration) as our proposal for a distance between theories. As we will see,  it reproduces the moduli space distance in the absence of a scalar potential.  However, our notion ends up having some intriguing properties when the potential is included. For instance,  the distance becomes energy dependent, as it depends on the energy of the path in configuration space connecting the two theories. Hence, it approximates the moduli space distance in the UV (when the potential becomes negligible) and as we decrease the energy, the potential becomes important.  So in a sense it is a notion of distance provided the available energy at a starting point. 
Therefore it is not a distance in the mathematical sense, since it will not satisfy standard properties like the triangular inequality as it also depends on the initial energy. In some sense, it is more appropriate to regard it as a ``cost function'', since it quantifies the difficulty in reaching faraway regions in moduli space given an initial finite resource (energy), but we refer to it as distance throughout this manuscript.

In the case of supersymmetric AdS vacua separated by a field-theory domain wall, our notion of distance also reproduces the expected AdS distance in \cite{lust2019ads}. We view this as a nice consistency check of our proposal. Assuming a relationship $m\sim\sqrt{\Lambda}$ for the mass of the tower of particles that becomes light in the $\Lambda\rightarrow0$ limit, we can reconstruct a relationship between the distance and the mass of the tower. We are able to recover the expected exponent of the Distance Conjecture, only for the case when there is exactly one extra dimension decompactifying. This is consistent with top-down holographic examples because in the formalism that we develop here we restrict ourselves to scalar potentials and scalar moduli spaces, and ignore e.g. effects of higher-dimensional p-form fields and other degrees of freedom. With these restrictions, indeed, the only possible flux compactification we can access is when there is a 1-form flux for an axion, and indeed, making this large leads one to decompactification of a single extra dimensions. Understanding how cases like e.g. $AdS_5\times S^5$, which involve 5-form flux, enter in this picture, is left for future work.

The organization of this paper is as follows:  In section \ref{nongrav} we discuss the domain wall tension as a measure of moduli space distance. We then study the distance between vacua with zero vacuum energy, leading to Minkowski backgrounds.  In section \ref{grav} we study the extensions of this to the case where the vacuum energy changes between vacua, and in this case we need to take into account the gravitational backreaction.  In section \ref{examples}, we offer a number of examples. In particular, we show why the supersymmetric case offers a particularly nice example where the distance can be readily computed in the case of minimal initial energy.  In section \ref{discussion}, we end the paper with some concluding thoughts.

\section{Proposal for a distance in a fixed background}
\label{nongrav}
\subsection{Domain wall tension as a moduli space distance}
To begin with, let us ignore the effects of gravity, and consider two different vacua with zero vacuum energy in a d-dimensional effective theory with a fixed background. For simplicity, let us restrict ourselves to the case of  homogeneous configurations, such that the two vacua only differ by the vacuum expectation values of some scalar fields.  The goal is to find a notion of distance between these two vacua.  In the case in which no potential barrier exists, namely the scalar fields are massless, there is a well-defined notion of distance measured by the field metric of the kinetic term of the scalars. This is known as the moduli space distance and it is given by
\beq
\Delta\phi=\int\sqrt{g_{ij}d\phi^id\phi^j} \ , \quad \text{with} \ \mathcal{L}\supset \frac12 g_{ij}\partial\phi^i\partial\phi^j+\dots.
\eeq
We now explain how this moduli space distance can be recovered from the tension of the domain wall connecting the two vacua, so that we can use the domain wall tension to define a generalised notion of distance in the presence of a potential in the next subsection.

For the sake of simplicity, let us first focus on the case of a single massless scalar field which is canonically normalised (i.e. $g_{ii}=1$). Consider a domain wall connecting the two vacua  characterized by the values $\phi_i$ and $\phi_f$. The tension of the domain wall is given by the minimal energy (per unit volume) of the path field configuration $\phi(\tau)$ interpolating between the two vacua, where $\tau$ is a spatial coordinate. 

Since there is no potential, the solution to the equation of motion is simply a trajectory of constant speed
\beq
\frac12\left(\frac{d\phi}{d\tau}\right)^2=\rho_E\quad \rightarrow\quad \phi(\tau)=\phi_0+\sqrt{2\rho_E}\tau,
\label{fji}\eeq
with $\sqrt{2\rho_E}=\Delta\phi/\Delta\tau$ and $\Delta\phi=\phi_f-\phi_i$. Hence, the tension of the solution interpolating between the two EFTs reads
\beq 
T=\int \left(\frac{d\phi}{d\tau}\right)^2d\tau=\int 2\rho_E d\tau=\frac{(\Delta\phi)^2}{\Delta\tau}.
\eeq
As a final step to get the domain wall tension, we need to minimize over all possible trajectories with different values of $\Delta \tau$, which implies that the tension vanishes when connecting two vacua in the same moduli space. This is because the minimal value occurs when the domain wall becomes infinitely thick ($\Delta\tau\rightarrow \infty$), i.e. when it spreads over the entire space, forcing the tension to vanish. However, we can recover the moduli space distance from the following \emph{normalized} tension 
\beq
\Delta\equiv\frac{T}{\sqrt{2\rho_E}}=\sqrt{2\rho_E}\Delta\tau=\Delta\phi,
\eeq
this normalized tension will be our starting point to define a generalized notion of distance in more complicated scenarios in the following. At the moment, dividing by $\sqrt{2\rho_E}$ can simply be seen as a rescaling of the distance which is required both to get the correct dimensions (i.e. a distance with units of energy) and to reproduce the moduli space distance in the absence of the potential.

We also notice that, strictly speaking, the solution \eqref{fji} does not interpolate between the two vacua (in the sense of Lorentz invariant states where the scalar field is homogeneous in space and time), but rather between the field \emph{values} $\phi_i$ and $\phi_f$. This feature will be present in the rest of the paper, when scalar potential and the effects of gravity are included, and sets apart our approach from e.g. the usual solitonic Coleman-De Luccia domain wall \cite{coleman1977fate, coleman1980gravitational}, where the vacuum is approached asymptotically for large values of $\tau$. By demanding only that the solution interpolates between the two vacua only at a particular value of the $\tau$ coordinate, we have a much more flexible framework, and can define the generalized distance in situations where the soliton and Coleman-De Luccia solutions do not exist (such as when interpolating between two vacua with different asymptotic values of the scalar potential, as we will see momentarily).

\subsection{Distance in the presence of a scalar potential}\label{sec:Mapertuis}

\begin{figure}[h]
	\centering
	\includegraphics[width=0.5\linewidth]{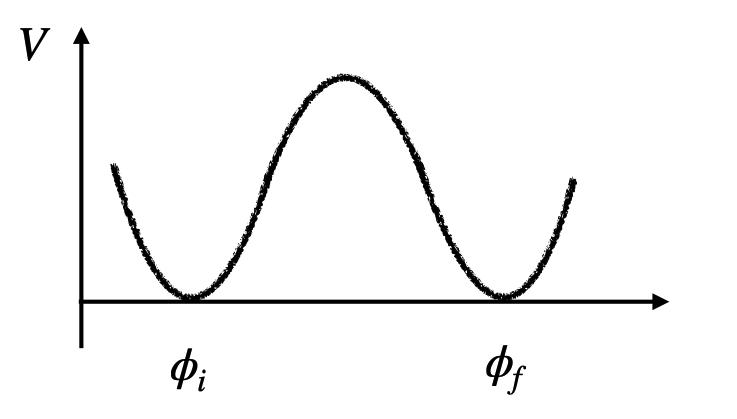}
	\caption{Two Minkowski vacua, separated by a barrier much smaller than the Planck mass.}
	\label{potential1}
\end{figure}

Let us now move beyond moduli spaces by adding a scalar potential to the theory. To start with, let us still ignore gravity and consider two field values $\phi_i$ and $\phi_f$ where the potential still vanishes (so $V(\phi_i)=V(\phi_f)=0$), separated now by a potential barrier as illustrated in Figure \ref{potential1}. This is the usual setup where one can construct a one-dimensional soliton domain wall solution \cite{rajaraman1982solitons} that interpolates between the vacua, since they have the same energy. 

The tension of the domain wall separating the two vacua is given by
\beq
T=\int  \left(\frac{d\phi}{d\tau}\right)^2 d\tau,
\eeq
and the scalar field configuration satisfies the equations of motion
\beq
\frac{d^2\phi}{d\tau^2}=V',
\label{eq1}
\eeq
subject to the following boundary conditions 
\beq
\phi(-\infty)=\phi_i\ ,\ \phi(\infty)=\phi_f \ ,\ \left.\frac{d\phi}{d\tau}\right|_{\pm\infty}=0\ .
\label{bc}
\eeq
In general, when $V(\phi_i)\neq V(\phi_f)$, the minimization problem \eqref{eq1} does not have a solution with boundary conditions \eqref{bc}; however, as explained at the end of the previous subsection, in this case we relax \eqref{bc} to demand only that $\phi_i$ and $\phi_f$ are attained for some values of $\tau$. In general, this yields the following domain wall tension
\beq
\label{Tthin}
T=\int  2(\rho_E+V(\phi))d\tau=\int_{\phi_i}^{\phi_f}\sqrt{2(\rho_E+V(\phi))}d\phi,
\eeq
where we have used that the solution to \eqref{eq1} is given by
\beq
\frac12\left(\frac{d\phi}{d\tau}\right)^2=V+\rho_E.
\eeq
The integration constant $\rho_E$ must vanish in this case where $V(\phi_i)=V(\phi_f)=0$ (or more generally, be equal to the minimal vacuum energy $V_{min}$ at the endpoints, $\rho_E=-V(\phi_{1,2})$) in order to satisfy the boundary conditions \eqref{bc} and asymptote to the two vacua. This recovers the usual domain wall tension of Coleman de Luccia \cite{coleman1977fate,coleman1980gravitational}.

Notice that the above tension is equivalent to the on-shell euclidean action of the path of minimal action in configuration space that interpolates between $\phi_{1}$ and $\phi_{2}$ at a fixed energy $\rho_E$. Interestingly, this variational principle is known in classical mechanics as the Maupertuis principle. It differs from the Hamilton's principle in that it finds the trajectory of minimal action by varying over time for a fixed energy, instead of varying over energies for a fixed time duration (as in Hamilton's). Thus, the Maupertuis functional $S_M$ can be written as the Legendre transform of Hamilton's action: $S_H=S_M-\rho_ET$, where $T$ is the time duration of the trajectory. For instance, consider the euclidean action for a single scalar in one dimension (i.e., on classical mechanics),\beq
S_M=\int d\tau( \frac12\dot{\phi}^2+V(\phi))+\int d\tau \rho_E\ .
\eeq
This yields the following equations of motion at fixed energy $\rho_E$
\beq
\label{eomdecouple}\frac12\dot{\phi}^2-V(\phi)=\rho_E\ , \quad \ddot{\phi}=V'(\phi),
\eeq
where $\dot{}=\partial_{\tau}$, and  $'=\partial_{\phi}$. The above equations are solved by $\dot{\phi}=\sqrt{2(\rho_E+V)}$. Plugging this into the action we indeed get
\beq
S_M=\int d\tau \dot{\phi}^2=\int d\phi \,\sqrt{2(\rho_E+V)}\ ,
\label{acs}
\eeq
reproducing the domain wall tension when the euclidean energy $\rho_E=-V_{min}=0$.

Following the reasoning of the previous section, we propose the distance between two theories that only differ by the vev of one scalar field as 
\beq
\label{defD2}
\Delta(\rho_E)= \frac{1}{\sqrt{\rho_E}}\int_{\phi_{i}}^{\phi_{f}}  d\phi \,\sqrt{\rho_E+V}, 
\eeq
where we have added the normalization factor $\sqrt{2\rho_E}$ to recover the moduli space distance in the absence of the potential.
This can be easily generalized to the case of multiple scalar fields with field metric $g_{ij}$, yielding
\beq
\label{distdef}
\Delta(\rho_E)= \frac{1}{\sqrt{\rho_E}}\int_{\phi_{i}}^{\phi_{f}}  \sqrt{g_{ij}(\rho_E+V)d\phi^id\phi^j}\ ,
\eeq
where we integrate over the path in configuration space that minimizes the euclidean action. This formulation in terms of the Maupertuis action makes manifest that our proposed notion of distance comes from a well-defined variational principle, so it will satisfy the required properties to behave as a geodesic distance in metric spaces. In fact, notice that \eqref{distdef} looks very similar to the geodesic distance in field space, except for the fact that the metric gets multiplied by a conformal factor equal to $\sqrt{\rho_E+V}$. Interestingly, this new metric $ J_{ij}=2(\rho_E+V)g_{ij}$ is known as the \textit{Jacobi metric} in classical mechanics, and it is well known that geodesics of this metric are equivalent to the trajectories of constant energy $\rho_E$ solving the equations of motion\footnote{In other words, the problem of finding constant-energy trajectories in a field theory with a scalar potential is equivalent to finding free geodesics of the Jacobi metric.} \cite{goldstein1980classical}.

Since the Jacobi metric is a metric in the traditional sense, the above definition satisfies all required properties to behave as a proper distance:
\begin{itemize}
\item It is positive definite. As long as the kinetic energy is quadratic in the velocity components, the Maupertuis action is always positive for all trajectories:
\beq
\Delta=\int \sqrt{1+\frac{V}{\rho_E}}d\phi= \frac{1}{\sqrt {2\rho_E}}\int \dot \phi \,d\phi= \frac{1}{\sqrt{2\rho_E}}\int \dot\phi^2d\tau\geq 0.
\eeq
\item For a fixed $\rho_E$, it is symmetric so that $\Delta(\phi_1,\phi_2)=\Delta(\phi_2,\phi_1)$ and  $\Delta(\phi_1,\phi_1)=0 $.
\item It satisfies the triangular inequality $\Delta(\phi_1,\phi_2) + \Delta(\phi_2,\phi_3) \geq \Delta(\phi_1,\phi_3)$. We are evaluating the on-shell action along the trajectory that minimizes the action, so it satisfies the triangular inequality by default as long as we are comparing trajectories with the same euclidean energy $\rho_E$.
\end{itemize}

Recall that the above distance is only related to the CdL domain wall tension when $\rho_E=-V_{\text min}$ (namely, $\rho_E=0$ for Minkowski vacua). However, it behaves as a geodesic distance for any value of $\rho_E$ due to the above properties. This opens up an interesting possibility: we can use the on-shell Euclidean Maupertuis action to define a generalized notion of distance $\Delta(\rho_E)$ between theories which is $\rho_E$-dependent. This is interesting for the following reason. First of all, in field theory, it allows us to define a distance between theories even in the case in which they cannot be connected by a planar domain wall (e.g. because they have different vacuum energies). 
If $\rho_E\gg V$, the distance \eqref{distdef} reproduces the moduli space distance, i.e. $\Delta(\rho_E\gg V)\approx \Delta\phi$, while if $\rho_E\sim V$, the effect of the potential barrier becomes more important and the distance between the theories increases, 
until  it reaches its maximum value given by 
the normalized tension of the CdL domain wall. This makes sense intuitively, as the potential barrier makes it more difficult to connect the two theories.
In the case of connecting two Minkowski vacua, the distance diverges in the limit $\rho_E\rightarrow 0$. Hence, we get a notion of distance that interpolates between the moduli space distance and the normalized CdL domain wall tension. Furthermore, it reproduces the field distance in the moduli space in the absence of a scalar potential, independely of the value of $\rho_E$. The cobordism conjecture (i.e. the existence of a finite energy domain wall interpolating between any two vacua of the same space-time dimension) translates then to the statement that any two theories should be at finite distance for non-vanishing $\rho_E$. Notice that $\rho_E$ only measures the energy density of the Euclidean configuration. In Lorentzian signature, it corresponds to a non-vanishing spatial derivative of the scalar field.

To summarize, in the context of field theory (i.e. ignoring gravitational backreaction) we have a perfectly reasonable one-parameter notion of distance $\Delta(\rho_E)$, where the reference energy scale $\rho_E$ provides a notion of ``scale dependence'' of the distance. In the next section we will see, however, that this is significantly complicated by gravitational effects: since there is no notion of energy in general relativity for general spacetimes, the scale $\rho_E$ loses its meaning as a conserved quantity, and indeed, it can change with $\tau$.

\section{Generalization of distance including gravitational effects}
\label{grav}

\begin{figure}[h]
	\centering
	\includegraphics[width=0.6\linewidth]{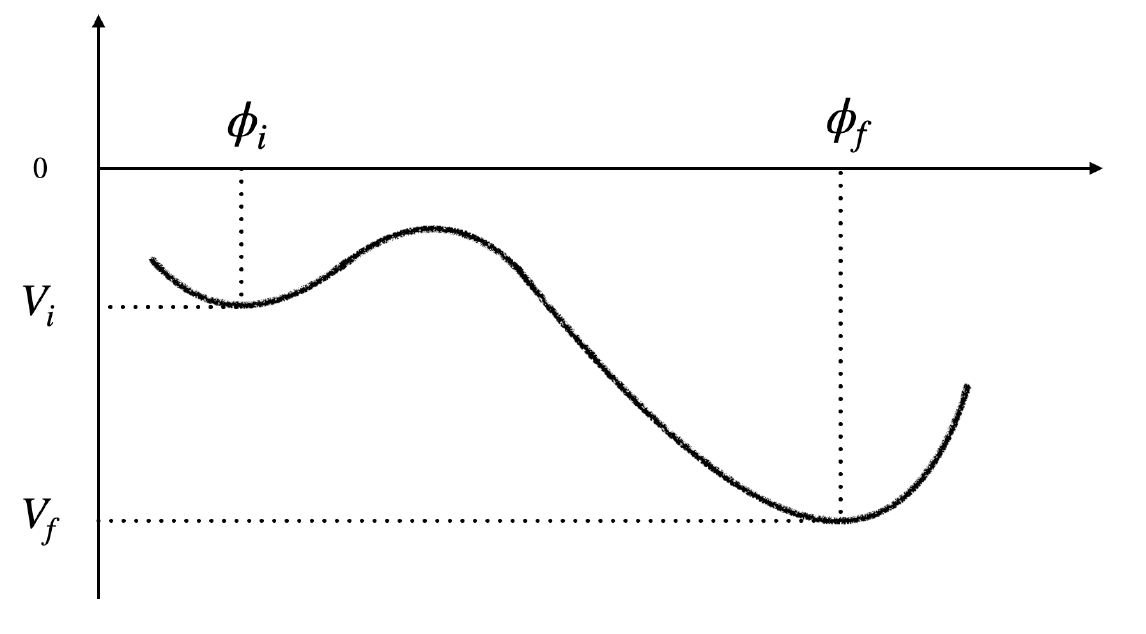}
	\caption{The scalar potential, $V(\phi)$, has minima at $\phi_i$, and $\phi_f$ which are the false and true vacua respectively.}
	\label{potential2}
\end{figure}

In this section we investigate the case in which the background is also dynamical, to account for gravitational effects. Consider a d-dimensional scalar field theory weakly coupled to gravity with a scalar potential, $V(\phi)$. Assume the scalar potential has two minima at $\phi_i$, and $\phi_f$ with $V_{f}\leq V_{i}$, where the minima are allowed to have different vacuum energies (see figure \ref{potential2}). For simplicity, let us assume that we want to compute the distance between the minima of this scalar potential  with negative energy, i.e, between two AdS vacua.   We will try to construct the configuration that we discussed in the previous subsection, this time taking into account an Einstein-Hilbert piece of the action. As we will see, general relativity implies that the two vacua are no longer connected by a configuration with constant euclidean energy. Hence, the conservation of energy can no longer be assumed, implying that we cannot use the Maupertuis variational principles of the previous section.   
  The metric ansatz for a domain wall is given by
\be \label{DWmetric}
ds^2=d\tau^2+a^2(\tau)d\sigma_\Sigma^2, \ee 
where $d\sigma_\Sigma^2$ is the metric on the space $\Sigma_{d-1}$ spanned by the domain wall, which we assume for the moment to be maximally symmetric and flat, while $\tau$ is the spatial coordinate transverse to the domain wall.  

We will also define
\be \Lambda(\phi)\equiv V(\phi)-\frac{1}{2}\dot{\phi}^2.\label{Ldef}\ee
Notice that $\Lambda(\phi)$ has the same definiton as $-\rho_E$ in the previous subsection; however, 
in the presence of gravity, the Euclidean equations of motion that we need to satisfy read
\be \label{scalarfieldEOM2} \ddot{\phi}+(d-1) \frac{\dot{a}}{a} \dot{\phi}= V'(\phi), \ee
\be \label{Einsteinequ2} \frac{\dot{a}^2}{a^2}=\frac{2\kappa^{\frac{d-2}{2}}}{(d-1)(d-2)}\left(\frac{1}{2}\dot{\phi}^2-V(\phi)\right), \ee
where $\kappa=8\pi G$. In terms of $\Lambda(\phi)$, the above two equations of motion can be simply recast as
\be \label{neweom2} \Lambda'^2=\alpha\kappa^{\frac{(d-2)}{2}}\Lambda(\Lambda-V),\quad \text{ with }\quad  \alpha=4 \frac{d-1}{d-2}\ , \ee
we see immediately that $\Lambda(\phi)$ is not a conserved quantity, as advertised. 

The equation of motion is first order, and so it needs a boundary condition. We will take

\beq
\label{initialL}
\Lambda(\phi_i)=V(\phi_i)-\left.\frac12 \dot \phi^2\right|_{\phi=\phi_i}\equiv \Lambda_i,
\eeq
for some constant $\Lambda_i$. This parameter is the \emph{initial} value of $\Lambda$, and it plays the same role as the $\rho_E$ in the theory without gravity.   This leads to a unique solution for $\Lambda (\phi)$ up to a sign ambiguity in choosing the square root in equation \eqref{neweom2} which will be explained below.

Once the equations of motion are resolved, the question remains of determining which functional of the on-shell solution we should use as a notion of distance. As illustrated in Section \ref{nongrav}, in the case without gravity we divided the domain wall tension by $\sqrt{2\rho_E}$ to ensure agreement with the moduli space metric. This is a property we would like to preserve in the presence of gravity, but now $\rho_E=-\Lambda$ depends on $\phi$. We would also like to recover the formula of the previous subsection in the limit where $\kappa\rightarrow0$. The simplest generalization of the flat space formula with these properties is to define our scale-dependent distance between the theories with scalar vevs $\phi_i$ and $\phi_f$ (including gravitational effects) as
\ba \label{gravdistance} \Delta(\Lambda_i)\equiv  \int_{\phi_i}^{\phi_f}\sqrt{1-\frac{V(\phi)}{\Lambda(\phi)}}d\phi, \label{distancedef2}\ea
for $\Lambda$ satisfying \eqref{neweom2} subject to the boundary condition \eqref{initialL}.

The outcome is that we obtain a scale-dependent distance, which depends on the initial value of $\Lambda_i$, such that it again reproduces the moduli space distance in the limit $\Lambda_i\gg V$. Also, in the gravitational decoupling limit, $|\Lambda|, |V|\ll \kappa^{-\sfrac{d}{2}}$, we recover the Maupertuis distance. In this limit, we have $\kappa\rightarrow 0$, and $ \sfrac{\dot{a}}{a}\rightarrow 0$ such that
\be \label{decouple} \frac{(d-1)(d-2)}{2}\kappa^{-\frac{d-2}{2}}\left(\frac{\dot{a}}{a}\right)^2\equiv \rho_E,\ee
remains constant. Upon taking the decoupling limit, the equations of motion reduce to \eqref{eomdecouple}, and we recover the Jacobi metric. We can interpret \eqref{gravdistance} as a recipe to compute the gravitationally backreacted distance by first chopping up the problem into many infinitesimal field displacements where gravity should not matter, using \eqref{distdef} in each of these, and add up the results  such that $\Delta=\int \Delta_{\text{flat space}}\,d\phi$. As we explain in more detail later, this varying $\Lambda(\phi)$ has both advantages and disadvantages. On the one hand, we recover the AdS distance proposed in \cite{lust2019ads}. On the other, we will lose the connection with a metric distance (satisfying the triangular inequality) in certain cases.

The initial value $\Lambda_i$ is bounded above, since the initial Euclidean kinetic energy must be large enough to guarantee
\ba
\label{bcL}
	\Lambda - V \leq 0, \quad  \text{for } \Lambda \leq 0\ ,
\ea
so that the distance is well-defined. Otherwise, there is no physical solution connecting the two theories. To guarantee monotonicity, the condition is that $\Lambda - V < 0$ and $\Lambda < 0$ between the initial and final vacua. This ensures that the right-hand side of \eqref{neweom2} does not cross zero in between $\phi_i$ and $\phi_f$, so $\Lambda(\phi)$ becomes a monotonic function of the scalar field. For a generic potential with the convention $V_f\leq V_i$, a sufficient (but not necessary) condition that guarantees \eqref{bcL} (and monotonicity) is
\be
	\Lambda_i < \min\{V_{\text{min}}, 0\},
\ee
where $V_{\text{min}}$ represents the minimum value of the scalar potential in the interval defined by $\phi_i$ and $\phi_f$. This condition can be relaxed depending on the shape of the scalar potential. 

Interestingly, this notion of distance is also related to a suitable generalized notion of domain wall tension beyond the thin-wall approximation, valid also in the gravitiational context \cite{Espinosa:2023tuj}. The tension of the (possibly thick) domain wall solution is given by 
\beq
T=\int_{\phi_i}^{\tilde \phi_f}\sqrt{2(V(\phi)-\Lambda(\phi))}d\phi,
\label{Tgrav}
\eeq
where $\Lambda(\phi)$ satisfies\footnote{This function $\Lambda(\phi)$ is known in \cite{Espinosa:2018hue,Espinosa:2018voj,Espinosa:2023tuj} as the tunneling potential $V_t(\phi)$.} precisely \eqref{neweom2} and the field value $\tilde{\phi}_2$, known in the literature as the ``tunneling exit point'', is defined by the condition $\Lambda(\tilde \phi_2)=V(\tilde \phi_2)$ --
namely, vanishing euclidean kinetic energy.
Strictly speaking, this implies that $\tilde{\phi}_2$ in \eqref{Tgrav} can be smaller than the value of the minimum $\phi_2$ (if it is larger, then the domain wall solution does not exist). In the thin wall approximation, one has $V_f\sim V_i$ so that $\Lambda(\phi)$ is approximately constant and $\tilde \phi_2\approx \phi_2$, so that \eqref{Tgrav} reproduces the thin wall tension result \eqref{Tthin}.

Finally, note that the differential equation \eqref{neweom2} has two possible solutions that differ on the sign of $\Lambda'$,
\be
	\Lambda' =\pm \sqrt{\alpha\kappa^{\frac{(d-2)}{2}}\Lambda(\Lambda - V)}. \label{lambdaprime}
\ee
Since $\Lambda'=\kappa^{\sfrac{1}{2}}(d-1)\frac{\dot{a}}{a}\dot{\phi}$, the choice of sign is correlated to providing a gravitational boost or friction to the accelaration of the scalar field. 
Since $\Lambda'$ is a monotonic function, the sign will be well defined along the entire path, yielding a priori two possible values for the distance. To have a unique well-define notion of distance, we choose the sign in correlation to the sign of the variation of the scalar potential, namely
\be\label{sign} sign(\Lambda')=sign(V_f-V_i),\ee
which corresponds to the boosting solution. For a fixed initial $\Lambda_i$, this actually corresponds to the larger of the two possible distances between the two vacua. This again guarantees that we can always find a solution, since the solution associated with the shorter distance does not always exist. For $V_f=V_i$, we again choose the larger notion of distance if the two choices differ.  When $V(\phi)=0$ the choice of sign does not matter, and we recover the expected result with either sign.

In Appendix \ref{multifield} we generalize this distance to multiple scalar fields, obtaining the following result as expected
\ba \Delta(\Lambda_i)\equiv  \int_{\phi_i}^{\phi_f}\sqrt{g_{ij}\left(1-\frac{V}{\Lambda}\right)d\phi^id\phi^j}, \label{distancedef3}\ea
with $\Lambda_i=V_i-\frac12 g_{ij}\dot\phi^i\dot\phi^j|_{\phi=\phi_i}$.

\subsection{Recovering the AdS Distance}
In this section we show that our proposal reproduces the AdS Distance proposed in \cite{lust2019ads} in the limit in which $\dot\phi$ approximately vanishes at the endpoints. Hence, our proposed distance interpolates between the moduli space distance when $\Lambda_i\gg V_i$ and the AdS Distance when $\Lambda_i\sim V_i$, at least in the cases where the AdS potential is driven by scalar fields. 

By plugging \eqref{lambdaprime} into \eqref{distancedef3},  we can write the distance simply in terms of variations of the euclidean energy density 
\be \label{2ndformdistance1} \Delta(\Lambda_i)= \frac{1}{\kappa^{\frac{d-2}{4}}\sqrt{\alpha}} \int_{\Lambda_i}^{\Lambda_f} \frac{d \Lambda}{\Lambda}, \ee
which becomes
\be  \label{logL}\Delta(\Lambda_i)= \frac{1}{\kappa^{\frac{d-2}{4}}\sqrt{\alpha}} \text{Log}\left(\frac{\Lambda_f}{\Lambda_i}\right) , \ee
so that it only depends on the value of $\Lambda$ at the endpoints. Recall that $\Lambda_i$ is an initial arbitrary condition, while $\Lambda_f$ gets fixed by solving the equation of motion \eqref{neweom2}. Whenever there exists a solution such that $\Lambda_f=V_f$ starting with $\Lambda_i=V_i$, one exactly recovers the AdS Distance $\Delta\sim  \log(V_f/V_i)$ from \eqref{logL}. This occurs whenever there exists a static planar domain wall between the two AdS vacua, as occurs in supersymmetric vacua connected by a BPS domain wall (see Section \ref{supersymmetric}).

Known supersymmetric AdS vacua arising from string theory do not exhibit scale separation between the AdS scale and the size of the extra dimensions, implying that there is a Kaluza-Klein tower of states scaling as $m\sim V_0^{1/2}$. Assuming the scaling $m\sim V^{\frac{1}{2}}$, we can derive the exponential rate of this tower in terms of the distance $\Delta$, thus obtaining 
\begin{equation} m\sim \Lambda^{1/2}\sim {\rm exp} (- {\kappa^{\frac{(d-2)}{4}}\sqrt \alpha\Delta/2})\sim {\rm exp}\left[-\kappa^{\frac{(d-2)}{4}}\sqrt\frac{d-1}{d-2}\Delta \right].\label{exp} \end{equation}
The exponential mass decay rate for a KK tower decompactifying from a lower $d$-dimensional vacuum to a higher $D$-dimensional vacuum is
$\sqrt \frac{D-2}{(d-2)(D-d)}$. Therefore, the exponential rate in \eqref{exp} coincides with the rate of a KK tower associated to the decompactification of one extra dimension. It is remarkable that using our definition of distance inspired by the domain wall tension we are getting a recognizable exponent for moduli space distances. Of course, it is not clear whether the exponential rates in the AdS case should match those associated to the moduli space distance. But given the coincidence, one may wonder why we are getting an exponent which only corresponds to one extra dimension opening up.  One explanation may be that since we are not studying the situation with fluxes in AdS and only with scalar, and since an axionic scalar can only give rise to flux with one higher dimension, this could be the reason we only are getting this situation.

\subsection{Properties and drawbacks of the scale-dependent distance}

This proposal for distance has a couple of interesting properties:
\begin{itemize}
\item It boils down to moduli distance, $\Delta_{\Lambda_i}=\int d \phi$, for $V\rightarrow 0$. 
\item It is positive definite 
\begin{align}\Delta_{\Lambda_i}= \int\frac{\dot{\phi}^2}{\sqrt{-2\Lambda}}d\tau\geq0.\end{align}
\item It depends on the initial scale $\Lambda_i$, such that it interpolates between the moduli space distance for $\Lambda_i\gg V_i$ and the AdS Distance for $\Lambda_i\sim V_i$.
\end{itemize}

However, this latter scale-dependent property also comes with some drawbacks. The distance is not symmetric anymore, since it depends on the initial value of $\Lambda_i$. Therefore, we need to adjust the initial conditions suitably to obtain a symmetric solution. In other words, we have $\Delta_{\Lambda^B_i}(B,A)=\Delta_{\Lambda^A_i}(A,B)$, which can be checked to be consistent with our sign choice for $\Lambda'$ in \eqref{sign}. Related to this, the triangular inequality is not guaranteed in general, although it holds whenever we have a solution with $V_i=\Lambda_i$ at both endpoints (like in the supersymmetric case in section \ref{supersymmetric}). In that case, the distance only depends on the initial and final values of the potential, and it therefore satisfies triangular equality (as in one-dimensional spaces). However, in more generic cases, our proposal does not behave as a distance in a metric space. In this sense, it should be better understood as a kind of cost function, which is scale-dependent. In the next section, we comment on other possible generalizations to try to overcome this issue, although it is not clear to us to what extent this failure of behaving as a metric distance is really a problem or a necessary feature. Intuitively, it is not unreasonable that the notion of distance loses its symmetric property, since roughly speaking, it costs more energy to go up a potential than going down, as any mountain hiker will confirm.

\subsection{Other possible generalizations}
The generalized distance we just proposed was relatively ad hoc, motivated mostly by the connection to the AdS distance and domain wall tension, and it did not come from an action principle in the gravitational case. One could try instead to directly use the gravitational generalization of the Maupertuis action or Jacobi metric that we used in the field theory case. The Jacobi metric for the time-dependent system at hand is given by \cite{cariglia2018cosmological}
\be \label{grav-Eis}
J_{ij}{d\phi}^i{d\phi}^j =2 \kappa^{\sfrac{d}{2}} a^{d-1} V(\phi) (-(d-1)(d-2) \kappa^{-\frac{d-2}{2}} a^{d-3}{da}^2 + a^{d-1}{d\phi}^2),
\ee
and the line element in field space is
\be
dl = \sqrt{J_{ij}{d\phi}^i{d\phi}^j}.
\ee
Notice that, unlike in the field theory case, this metric is not positive definite. This happens whenever one has a system with constraints; in the case of gravity, not every initial condition is valid, since they must satisfy the Hamiltonian constraint of GR \cite{wald2010general}. In other words, for the above metric to be equivalent to the solution of Einstein's equations,  the physical time is defined via
\be \frac{(d-1)(d-2)}{2}\kappa^{-\frac{(d-2)}{2}}\frac{\dot{a}^2}{a^2}=\frac{1}{2}\dot{\phi}^2-V(\phi). \ee
 It turns out that, when this constraint is taken into account and the on-shell action is evaluated, the dynamical component of gravity cancels the contribution of matter fields to the distance, except for the scalar potential term. This line element vanishes when $ V = 0 $. Therefore, although it is a good metric, it does not straightforwardly lead to a good measure of distance. Extracting a notion of distance from this metric is left for future work.  Our proposed distance of the previous subsections corresponds to the piece of the above Maupertuis action associated only to the scalar field, which recovers the moduli space distance when $V=0$ at the cost of losing the triangular inequality (although we have triangular equality for supersymmetric configurations). We also note that the above metric has been derived using a technique called \textit{the Eisenhart lift}, which leads to a metric in the higher-dimensional field space resulting from adding a dummy field (degree of freedom). To obtain a metric in the lower-dimensional field space from the Eisenhart line element, one should project the Eisenhart line element onto the energy constraint surfaces, which indeed leads to the proposed Jacobi metric, \eqref{distdef}, in the energy-constant case and \eqref{grav-Eis} in the gravitational case.\footnote{Note that the on-shell Eisenhart action does not yield a metric in the lower-dimensional field space.}

A related, perhaps more physical proposal starts by trying to construct not an Euclidean solution in $d$-dimensions, but instead, a valid \emph{initial} condition for general relativity, and try to minimize its mass. More specifically, we could look at the ADM mass of a generic bubble of radius $R$ (meaning it bounds a sphere of area $4\pi R^2$ interpolating between the two geometries), and find the field configuration that minimizes $M_{\text{ADM}}$. Details are provided in Appendix \ref{app:B}. The energy functional that must be minimized resembles the Maupertuis euclidean action of Section \ref{sec:Mapertuis}, although the equations of motion for the scalars are evaluated in one dimension less, since we are only looking for a solution at a initial time slice.  However, we find that in general (e.g. when the potential is too steep) an event horizon may form, in which case the formalism stops being valid. On top of this problem, we face the difficulty that the ADM mass is not universally positive in general relativity, and only in restricted circumstances (e.g. whenever a positive energy theorem applies) this will happen. Since the effective tension could happen to be negative (as e.g. for orientifolds, or in some sense for bubbles of nothing \cite{Friedrich:2023tid,Friedrich:2024aad}), it is not clear that this definition leads to the expected properties of a distance for metastable vacua. On the other hand, these difficulties might be ameliorated for supersymmetric, exactly stable vacua, where positivity of the ADM mass is expected to hold. Because of these difficulties, we have not explored this notion of distance any further, but the simple underlying result and direct physical interpretation in terms of a minimal mass are quite appealing. Furthermore, the connection to BPS domain walls and minima works in the same way as for the proposal described in the previous Section, since the notion of thin wall is similar in this case. We hope to return to this question in the future.

As a final comment, let us remark that we have only investigated the case in which the different field configurations can be connected by a planar domain wall. This is why we have focused on Section \ref{grav} in the case of AdS vacua, since they admit maximally symmetric flat slices in euclidean signature. However, if we want to have solutions interpolating between more general configurations (like e.g. de Sitter vacua, which are positively curved), we will have to add non-vanishing intrinsic curvature to the metric \eqref{DWmetric}. Upon doing so, the equation of motion for the scalar field \eqref{scalarfieldEOM2} does not change, while \eqref{Einsteinequ2} will be modified by a curvature term.\enlargethispage{-\baselineskip} If we keep our definition of \(\Lambda\) as minus the energy density of the scalar field, which accounts for both intrinsic and induced sources of curvature, the equation \eqref{neweom2} governing the dynamics of \(\Lambda\) is modified to $\Lambda'^2=\alpha \kappa^{\frac{(d-2)}{2}}(\Lambda + \frac{(d-1)(d-2)}{2} \frac{k}{a^2})(\Lambda-V)$, and consequently the logarithm \eqref{logL} is also modified by a term containing the effects of the intrinsic curvature\footnote{The distance would apparently get modified to 
$
\Delta = \frac{1}{\kappa^{\frac{d-2}{4}} \sqrt{\alpha}} 
\int \frac{d \Lambda}{\sqrt{\Lambda \left( \Lambda + \frac{(d-1)(d-2)}{2} \frac{k}{a^2} \right) }} 
$
where \(\frac{k}{a^2}\) is the intrinsic curvature.}. However, when written as in \eqref{gravdistance}, the formula for the distance remains unchanged. Since additional subtleties could appear in these cases, we leave the investigation of these curved solutions for future work.

\section{Examples}
\label{examples}
In this Section, we offer a couple of examples in the decoupling limit of gravity, $V\ll\kappa^{-\sfrac{d}{2}}$, using our metric of Section \ref{nongrav} (sections \ref{t3nongrav}, and \ref{CYnongrav}).
We also illustrate the generalized notion of distance explained in Section \ref{grav} with some examples, including the $\mathcal{N}=1$ supersymmetric example (sections \ref{supersymmetric}, \ref{constantgrav} and \ref{expgrav}).

\subsection{Distance between Supersymmetric Vacua}
\label{supersymmetric}
Consider a four-dimensional $\mathcal{N}=1$ supersymmetric theory 
\be S=\int dv_{\sigma} d\tau a^3\left[-3\frac{\dot{a}^2}{a^2}+K_{I\bar{J}}\dot{\phi}^I\dot{\bar{\phi}}^{\bar{J}}+V(\phi^I,\bar{\phi}^{\bar{I}})\right], \ee
with the scalar potential of the theory given by
\be V=e^{K}\left(K^{I\bar{J}}D_{I}WD_{\bar{J}}\overline{W}-3\,\left|W\right|^2\right). \ee
We set $\kappa=1$ throughout this section. Each superfield can contribute to supersymmetry breaking through a non-vanishing F-term, $ F_{I}\equiv D_{I}W= (\partial_I+\partial_I K) W $. We are interested in the solution interpolating between two supersymmetric AdS minima with $F_I^i = F_I^f = 0$. In between these two minima the supersymmetry is broken, $D_I W \neq 0$ (we can have a positive barrier for example). 

There exists a BPS domain wall solution connecting the vacua. The solution preserves $\mathcal{N}=\frac{1}{2}$ supersymmetry. The supersymmetry equations, $\delta\psi^{\mu} = \delta\chi^{I} = 0$, have the following solutions
\ba \frac{\dot{a}}{a}&=&\pm \left|e^{\sfrac{K}{2}}W\right|,\label{susysol1}\\
\dot{\phi}^I&=&\mp e^{\sfrac{K}{2}}K^{I\bar{J}}\frac{W\overline{D_J W}}{|W|},\label{susysol2} \ea
which also satisfy multi-field equations of motion \eqref{multi1}, and Einstein's equation \eqref{multi2} \cite{ceresole2006domain, cvetivc1992static}. 
 
We can reparametrize the path in the following way
\be d\varphi^2=2 K_{I\bar{J}}d\phi^{I}d\bar{\phi}^{\bar{J}}.\ee
For brevity, we define $\left|DW\right|^2 = K^{I\bar{J}}D_{I}WD_{\bar{J}}\overline{W}$. From \eqref{susysol2}, we get 
\be \frac{1}{2}\dot{\varphi}^2=K_{I\bar{J}}\dot{\phi}^I\dot{\bar{\phi}}^{\bar{J}}=e^K|DW|^2, \ee
so the kinetic term exactly cancels the F-term contribution to the scalar potential. Therefore, the total kinetic energy of the scalar fields is a measure of supersymmetry breaking in the original potential, and we have
\ba \Lambda=-3e^K|W|^2.\ea
At the initial and final supersymmetric vacua we have $\sfrac{1}{2}\dot{\varphi}^2=e^K|DW|^2=0$. Therefore we get
\ba \label{bc1}\Lambda(\phi^I_f)&=&V(\phi^I_f),\\
\label{bc2}\Lambda(\phi^I_i)&=&V(\phi^I_i).\ea
In this case, the value of $\Lambda$ is uniquely fixed at initial and final points. 

We can rewrite the solution for the trajectory in the following way
\ba &&\frac{d\phi^I}{d\varphi}=\pm\frac{K^{I\bar{J}}W\overline{D_{J}W}}{\sqrt{2}|W||DW|},\\
&&\frac{d\bar{\phi}^{\bar{I}}}{d\varphi}=\pm\frac{K^{\bar{I}J}\overline{W}D_{J}W}{\sqrt{2}|W||DW|}. \ea
We note that the trajectory aligns with the gradient of $\Lambda$
\ba &&\frac{\partial\Lambda}{\partial\phi^I}=-3e^K \overline{W}D_IW,\\
      &&\frac{\partial\Lambda}{\partial\bar{\phi}^{\bar{I}}}=-3e^K W\overline{D_IW}.\ea
As expected, we recover
\ba \frac{d\Lambda}{d\varphi}&=&\frac{\partial\Lambda}{\partial\phi^I} \frac{d\phi^I}{d\varphi}+\frac{\partial\Lambda}{\partial\bar{\phi}^{\bar{I}}}\frac{d\bar{\phi}^{\bar{I}}}{d\varphi},\\
&=&\mp 3\sqrt{2} e^K|W||DW|,\\
&=&\mp \sqrt{6\Lambda(\Lambda-V)}.\label{recoveom}\ea
As long as \( W,\, DW \neq 0 \) (which is the case in between the vacua), the sign does not flip and is determined according to equation \eqref{sign}. This allows us to switch between the square root and logarithm forms of the distance. The distance between two supersymmetric vacua is then given by
\begin{equation}
\label{distSUSY}
\Delta = \sqrt{1 - \frac{V}{\Lambda}} d\varphi = \frac{1}{\sqrt{6}} \text{Log} \left( \frac{\Lambda_f}{\Lambda_i} \right) = \frac{1}{\sqrt{6}} \text{Log} \left( \frac{V_f}{V_i} \right),
\end{equation}
where we have used the equation of motion \eqref{recoveom} in the first equality, and the boundary conditions \eqref{bc1}, and \eqref{bc2} in the second equality. We can rewrite the distance in terms of the tension of the BPS wall 
\begin{equation}\sigma = \frac{2}{\sqrt{3}} \left( |\Lambda_f|^{1/2} - |\Lambda_i|^{1/2} \right),\end{equation} in the following way
\begin{equation}
\Delta = \sqrt{\frac{2}{3}} \text{Log} \left( 1 + \frac{\sqrt{3}}{2} \frac{\sigma}{|V_i|^{1/2}} \right),
\end{equation}
which can be approximated by
\begin{equation}
\Delta \simeq \frac{1}{\sqrt{2}} \frac{\sigma}{|V_i|^{1/2}} + \dots.
\end{equation}
Therefore, the proposed distance reproduces the normalized tension of the BPS domain wall in the supersymmetric case, as long as one chooses the minimal value for the energy scale at the initial point (as given in \eqref{bc2}). In this case, the distance \eqref{distSUSY} satisfies the triangular inequality in a kind of trivial way, since it only depends on the value of the vacuum energy at the endpoints. This is specific to supersymmetric setups in which there exists a domain wall solution that starting with $\dot\varphi=0$ at the initial vacuum reaches again $\dot\varphi=0$ at the final vacuum.

Finally, we note that another typical property of a distance function is that it only vanishes when evaluated between a point and itself. This is not automatically true of our solutions. Indeed, there could be in principle $W=0$ solitonic domain walls between a theory an itself. Such domain walls were considered, and constructed, purely in the supergravity context in \cite{Cvetic:1992bf,Cvetic:1993xe} (see also the review \cite{Cvetic:1996vr}). For such a domain wall, the asymptotic value of $\Lambda_i$ would be equal to $\Lambda_f$, and the expression for the distance above would vanish. Interestingly, however, there are no known top-down examples where the phenomenon described in \cite{Cvetic:1992bf} takes place\footnote{We thank Mirjam Cveti\v{c} for pointing this out to us.}, which we take to be an encouraging sign for our proposal.

\subsection{Constant Scalar Potential; $V=V_0$ }
\label{constantgrav}
Take a d-dimensional theory with a constant scalar potential, $V=V_0\leq 0$. In this case the equation of motion is
\be \label{eomconst} \Lambda'=-\sqrt{\alpha\kappa^{\frac{(d-2)}{2}}\Lambda(\Lambda-V_0)} ,\ee
which can be easily solved. To start with, we set the initial condition $\Lambda(0)=V_0$, where we have assumed $\phi_i=0$ for simplicity. This initial condition implies that we start with the minimal possible energy at the initial point.  We find 
\be \Lambda=V_0\left[\text{cosh}\left(\frac{\sqrt{\alpha}}{2}\kappa^{\frac{(d-2)}{4}}\,\phi\right)\right]^2, \ee
so that the distance gets independent of the value of the potential
\ba \Delta&=& \int_{0}^{\phi}\text{tanh}\left(\frac{\sqrt{\alpha}}{2}\kappa^{\frac{(d-2)}{4}}\,\phi\right)\,d\phi,\\
&=&\frac{2}{\kappa^{\frac{(d-2)}{4}}\sqrt{\alpha}} \text{Log}\left(\text{cosh}\left(\frac{\sqrt{\alpha}}{2}\kappa^{\frac{(d-2)}{4}}\,\phi\right)\right). \ea
  On the other hand, if we choose the initial condition such that it includes a non-zero contribution $\rho_K$ from the kinetic energy at the initial point
 \be \Lambda(0)=-\rho_K+V_0,\ee
 with $\rho_K \gg V_0$, we get $\Lambda\simeq-\rho_K$ and we recover
 \be \Delta\simeq\int d\phi, \ee
 which is, as expected, the distance in moduli space.  We note that, although the scalar potential is constant, the euclidean energy $\Lambda(\phi)$ changes due to the gravitational backreaction. 

\subsection{Distance between two  $T^3$ compactifications connected by a massive deformation}
\label{t3nongrav}

Let us consider next the distance between two EFTs arising from a toroidal compactification of the same theory, such that they are no longer connected by a moduli space. For instance, we can consider a $T^3$ compactification with two different metrics related by a conformal transformation of the form
\beq
\tilde g_{\tilde T^3}=e^\varphi g_{T^3},
\eeq
where $ \varphi $ is a deformation which does not solve the Einstein's equations on  $T^3$. Hence, the  field $\varphi$ is not a modulus of this compactification. As an example, we can take $e^{2\varphi}=\phi(\lambda)(1+\cos(kx))$. The relevant part of the action is given by
\beq
S=\int_{T^3\times \mathbb{R}_\lambda} \mathcal{R}=\int_{T^3\times \mathbb{R}_\lambda}e^{2\varphi}\nabla_i\varphi\nabla^i\varphi+\dots = \int d\lambda \left((\partial_\lambda \phi)^2+\frac{k^2}4 \phi^2\right)+\dots,
\eeq
where we have already integrated over $T^3$ in the last step.
The resulting EFT is that of a massive scalar field in 0+1 euclidean dimensions. The on-shell action for the path solving the equation of motion at fixed energy $\rho_E$ is given by \eqref{acs} with potential $V=m^2\phi^2$ and $m^2=k^2/4$. We work in the decoupling limit, $V\ll\kappa^{-\sfrac{d}{2}}$, so that we can use \eqref{defD2} to obtain the distance between the two EFTs
\beq
\Delta(\rho_E)= \frac{1}{\sqrt{\rho_E}}\int_{\phi_{ i}}^{\phi_{ f}}  \sqrt{g_{ij}(\rho_E+V)d\phi^id\phi^j}=\int_{\phi_{i}}^{\phi_{ f}} \sqrt{1+\frac{m^2\phi^2}{\rho_E}}d\phi.
\eeq
Due to the potential, the result for the distance is energy dependent. If $\rho_E\gg V$, then we recover the moduli space distance again. However, as $\rho_E$ decreases, the distance between the two EFTs increases and diverges in the limit $\rho_E\rightarrow 0$. The interpretation of this IR divergence is that the potential obstructs the path, so that there is no domain wall solution that interpolates between the two theories unless we allow for some non-vanishing energy $\rho_E>0$.

\subsection{Exponential runaway potential}
\label{expgrav}
Consider a scalar potential of exponential form. This behavior is expected in the asymptotic region of the moduli space
\be
V(\phi) = -V_0 e^{c\,\kappa^{\frac{(d-2)}{4}}\phi},
\ee
where $c > 0$, and $V_0 > 0$. For  $0 < c < \sqrt{\alpha}$ with $\alpha$ given in \eqref{neweom2}, there is a solution for $\Lambda(\phi)$ for any arbitrary range in $\phi$
\be\Lambda(\phi) = -V_0\left(1 - \frac{c^2}{\alpha}\right)^{-1}e^{c\,\kappa^{\frac{(d-2)}{4}}\phi}.\ee
This is known as the \textit{attractor solution} in cosmology. The distance is then given by
\be\Delta = \frac{c}{\sqrt{\alpha}}\phi,\ee
which has a nice physical interpretation: For the attractor solution, when \(c = 0\), we have \(\Lambda = V = -V_0\), which corresponds to a fixed point with \(\dot{\phi} = 0\). In this case, the probe scalar field does not move, and the distance measure (maupertuis action density) is zero at each point. As \(c\) (the slope of the potential) increases, the kinetic energy becomes non-zero, and the (probed) distance increases. Finally, it matches the moduli distance for \(c = \sqrt{\alpha}\), where \textit{kination} occurs, and the distance is expected to match that of a free field. We note that the potentials obtained from string theory typically satisfy the above condition on \(c\).

\subsection{Distance between Calabi-Yau's connected by a conifold topological transition}
\label{CYnongrav}

We are now ready to try a slightly more involved but interesting case. One of the most interesting applications of our metric is that it can be used to provide a distance between different Calabi-Yau compactifications to Minkowski space. It is well known that different Calabi-Yau's can be often connected by a topological transition in which some cycles blow up and others blow down. From a physics perspective, if we have string theory compactified on these CYs', what happens is that a finite number of states coming from wrapping branes become massless at the finite distance singularity where the topological transition occurs.

As an illustrative example, let us consider a toy model for two Calabi-Yau's (CY$_1$ and CY$_2$) connected by a conifold-like topological transition.  In such cases, on one branch one field is massless and on the other side, a new field becomes massless.  In particular, we can view the relevant potential for the two fields $r, \rho$ by
\beq
V=\frac12 r^2 \rho^2,
\eeq
when one of the fields $r$ or $\rho$ is non-zero, it serves as mass for the other field.  

Let us now compute the distance between some regular point in the moduli space of CY$_1$ and another regular point in the moduli space of CY$_2$, such that the two points are in the vicinity of the conifold singularity. For this, let us work in the decoupling limit, $V\ll\kappa^{-\sfrac{d}{2}}$, so that we can use the notion of distance given in \eqref{distdef}. In Figure \ref{CYcontour} we represent (schematically) the two-field space parametrized by $r$ and $\rho$. 
Using \eqref{defD2}, the distance between the point $\phi_{i}=\{\rho=0,r=r_0\}$ and $\phi_{f}=\{\rho=\rho_0,r=0\}$ is given by
\beq
\Delta(\rho_E)= \int_{\phi_{i}}^{\phi_{f}} \sqrt{\left(1+\frac{r^2\rho^2}{2\rho_E}\right)\left(dr^2+ d\rho^2\right)}.
\eeq
The equations of motion are 
\ba &&\frac{1}{2}\left(\dot{r}^2+\dot{\rho}^2\right)-\frac{1}{2}r^2\rho^2=\rho_E,\\
&& \ddot{\rho}=r^2 \rho,\; \ddot{r}=\rho^2 r,
\ea
depending on the value of $\rho_E$, the path $r(\tau),\rho(\tau)$ minimizing the action will be different. Solutions for different values of $\rho_E$ are shown in Figure \ref{traj}. As the figure shows, for small values of the energy $\rho_E$, the trajectory is mainly controlled by the effect of the scalar potential, and the trajectories match the equipotential lines (the blue line). As \(\rho_E\) increases, the effect of the potential becomes more and more negligible, and the trajectory asymptotes towards the free trajectory, which is the straight line connecting the two points (the red line). The two extreme cases are given by:
\begin{figure}[t]
	\centering
	\begin{subfigure}[t]{0.5\textwidth}
		\centering
		\includegraphics[width=0.95\linewidth]{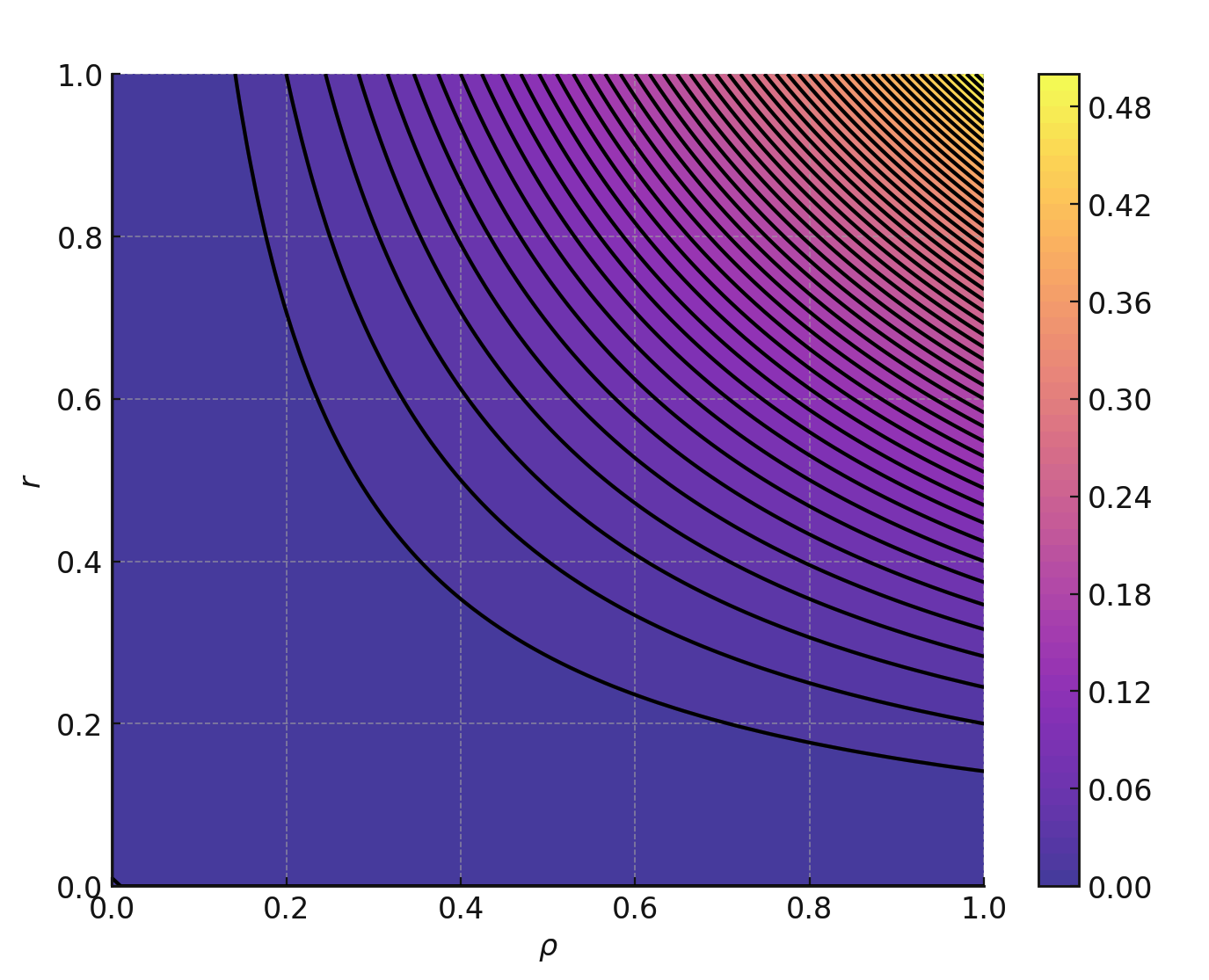}
           \caption{Contour plot of $V(\rho,r)$.}
		\label{CYcontour}
	\end{subfigure}%
	\begin{subfigure}[t]{0.5\textwidth}
		\centering
		\includegraphics[width=0.8\linewidth]{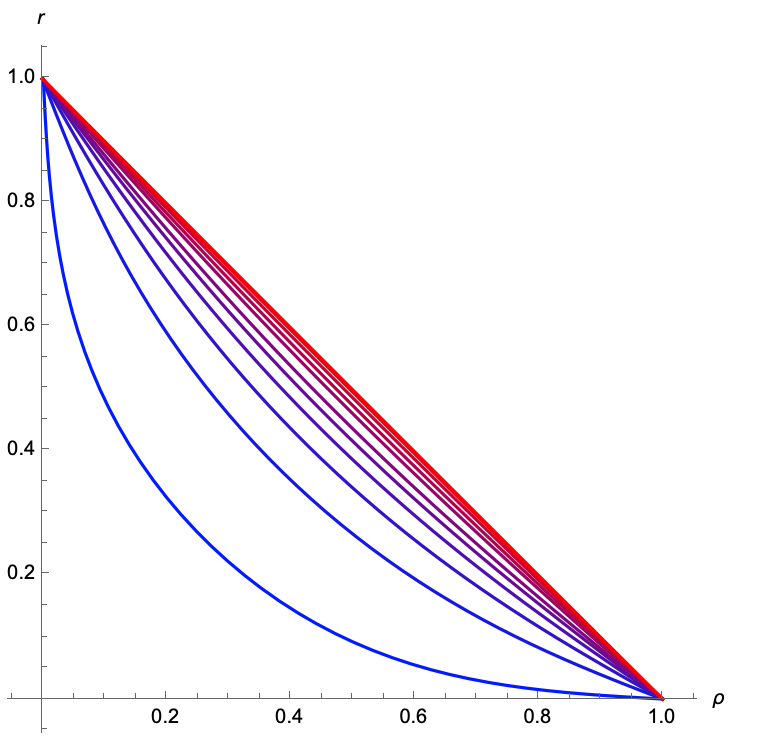}
         \caption{Constant energy trajectories interpolating between the points \(\{1,0\}\) and \(\{0,1\}\). The euclidean energy $\rho_E$ increases from blue to red.}
		\label{traj}
	\end{subfigure}
 \caption{Details of the scalar potential and distance for the conifold toy model discussed in the main text.}
\end{figure}
\begin{itemize}
\item As \(\rho_E \rightarrow 0\), the path approaches the valley of the potential (i.e. moving along the horizontal and vertical axes) and the distance behaves as $\Delta \simeq \frac{1}{\sqrt{\rho_E}} \int_{\phi_i}^{\phi_f} r \rho \sqrt{dr^2 + d\rho^2}$
for large \(r\) and \(\rho\), and approximates the flat one near \(r, \rho \sim 0\) evaluated on the moduli space geodesic.

\item As $\rho_E\rightarrow \infty$, it approaches the expected field space distance, $\Delta\simeq\int_{\phi_i}^{\phi_f}\sqrt{dr^2+d\rho^2}$ and moves along the straight line as if the potential would not exist.
\end{itemize}

The distance is a 
function of energy, that varies between the above two values.

\section{Discussion}
\label{discussion}
In this paper we have shown that the notion of moduli space distance in quantum field theory (when gravity is decoupled) can be naturally extended to the cases where there is a scalar potential on the field space using the Maupertuis action principle. This provides a new metric (the Jacobi metric) that takes into account the scalar potential and whose geodesics are the paths of fixed energy solving the euclidean equations of motion and therefore extremizing the euclidean Maupertuis action. Our notion of distance thus depends on an auxiliary energy scale, which plays the role of an ``RG scale'' for measuring distance. When this scale is much higher than any potential barrier, the distance reproduces the moduli space distance. On the other hand, when it gets comparable to the potential barrier, the distance gets larger and can even diverge. When this energy scale takes its minimal possible value, the distance approaches the value of the tension of the domain wall between vacua normalized by this energy scale.

The above notion of distance can also be generalized in the presence of gravity. Although in this case the resulting quantity does not obey in general triangular inequality or other expected properties of distance (and is, therefore, a ``generalized'' distance between field configurations at best), the quantity we found still quantifies a notion of ``distance to a given point'' in the space of vacua, providing a monotonic, positive function that increases away from our given reference point. The appealing feature of this generalized distance is that, as we vary the initial energy of the euclidean configuration connecting the two theories, it interpolates between the moduli space distance and the distance proposed in the AdS case for discrete families of vacua \cite{lust2019ads}, recovering this way the expected $\log V_0$ behavior with the vacuum
energy $V_0$. Moreover, it behaves as a distance in the mathematical sense when connecting supersymmetric AdS vacua, as it again gets directly related to the tension of the BPS domain wall between the vacua in that case.

We believe this notion has physical significance.  One important application of defining a distance on moduli spaces is to extend the Distance Conjecture to situations with a scalar potential. To do this, one does not obviously need the exact mathematical notion of distance (symmetric, positive, and satisfying triangle inequality); perhaps all that is needed is a notion of ``point at infinite distance'', as we provided. Since our notion of distance reproduces the moduli space distance and the AdS distance in certain limits, it seems well suited to provide such generalization of the Distance Conjecture. In this line, it is perhaps interesting that we get  the expected value of the exponential decay rate of the tower for one particular case corresponding to decompactifying one extra dimension in non-scale separated AdS vacua. It remains an open question how to completely generalize the Distance Conjecture using this notion of distance and how the mass of the tower should behave depending on the chosen value of the initial auxiliary energy scale. As emphasized in \cite{Stout:2022phm}, from a bottom-up perspective, the concept of distance is intended to capture the breakdown of the theory through the emergence of light tower of states as one approaches weak coupling limits or the decoupling limits of gravity. Our proposed cost function 
regains its metric-like properties in these limits, as noted in \eqref{decouple}. Furthermore, the cost function accurately describes the mass of the light tower in supersymmetric Anti-de Sitter spaces (also) within these limits, where it is expected to hold physical significance from a bottom-up standpoint. In this context, the cost function serves as  an effective measure that aligns with the bottom-up motivations of the Swampland Distance Conjecture.

Finally, we have also discussed other possible generalizations of the distance that come from variational principles including gravity but fail to reproduce the moduli space distance and/or the AdS Distance of \cite{lust2019ads}. Moreover, in this work we have focused on defining the distance between theories that only differ by the vev of some scalar fields, so that the domain wall can be smoothly described within the EFT using the effective potential for these scalars. It would be interesting to generalize this notion of distance to other field configurations, such as e.g. when we have gauge fields, fluxes, etc. In those cases, the domain walls might include singular defects that cannot be resolved within the EFT and one should probably consider the full euclidean action to define the distance. \\

\textbf{Acknowledgements}: We are grateful to Costas Bachas, Alek Bedroya, José Calderón Infante, Mirjam Cveti\v{c}, Jose Ram\'{o}n Espinosa, Jakob Moritz and Ignacio Ruiz for illuminating discussions and comments. The work of AM and CV is supported in part by a grant from the Simons Foundation (602883,CV), the DellaPietra Foundation, and by the NSF grant PHY-2013858. AM is thankful to the TH department of CERN, where most of the research on this project was carried out. M.M. is
supported by an Atraccion del Talento Fellowship 2022-
T1/TIC-23956 from Comunidad de Madrid. M.M. and I.V acknowledge the hospitality of  the
Department of Physics of Harvard University during the
different stages of this work, as well as the Simons Summer workshop and the Erwin Schrödinger International Institute for Mathematics and Physics of the University of Vienna. MM and IV thanks the Spanish Research Agency (Agencia Estatal de Investigación)
through the grants IFT Centro de Excelencia Severo
Ochoa CEX2020-001007-S and PID2021-123017NB-I00,
funded by MCIN/AEI/10.13039/501100011033 and by
ERDF A way of making Europe. The work of I.V. is also partly supported by the grant RYC2019-028512-I from the MCI (Spain) and the ERC Starting Grant QGuide-101042568 - StG 2021.

\appendix
\section{Generalization to Multiple Scalars}
\label{multifield}
Assume we have several scalars $\phi_k$ with $k=1,\dots, n$, the action is
\be S=\int dv_{\sigma} d\tau a^{d-1}\left[-\frac{1}{2} R+\frac{1}{2}g_{ij}(\phi^k)\dot{\phi}^i\dot{\phi}^j+V(\phi^k)\right],\ee 
 equations of motion for the scalar fields are as follows
\be\label{multi1}\ddot{\phi}^k+(d-1)\frac{\dot{a}}{a}\dot{\phi}^k+\Gamma^k_{ij}\dot{\phi}^i\dot{\phi}^j=g^{ki}V_{,\phi^i} ,\ee
 and Einstein's equation implies
 \be\label{multi2}  \Lambda(\tau)\equiv -\frac{(d-1)(d-2)}{2}\frac{\dot{a}^2}{a^2}=-\frac{1}{2}g_{ij}(\phi^k)\dot{\phi}^i\dot{\phi}^j+V(\phi^k).\ee
We can solve these $n+2$ equations (\eqref{multi1}, and \eqref{multi2}) for $\phi^k(\tau)$, $a(\tau)$, and $\Lambda(\tau)$. We define the field measuring length of the field space trajectory in the following way
\be d\varphi^2\equiv g_{ij}(\phi^k)d\phi^i d\phi^j.  \ee
We note that 
\be g_{ij}\frac{d\phi^i}{d\varphi}\frac{d\phi^j}{d\varphi}=1,\ee
 we can take a derivative with respect to $\varphi$ to get
  \be g_{ij}\frac{d^2\phi^i}{d\varphi^2}\frac{d\phi^j}{d\varphi}+g_{al}\Gamma^a_{ij}\frac{d\phi^i}{d\varphi}\frac{d\phi^j}{d\varphi}\frac{d\phi^l}{d\varphi}=0.\ee
   Upon projecting \eqref{multi1} on $g_{kl}\sfrac{d\phi^l}{d\varphi}$ (tangent to the path), and $\left(g_{kl}-g_{ki}g_{lj}\sfrac{d\phi^i}{d\varphi}\sfrac{d\phi^j}{d\varphi}\right)$ (orthogonal to the path) after using the above identities, we get
\ba \label{eomproj1} &&\ddot{\varphi}+(d-1)\frac{\dot{a}}{a}\dot{\varphi}=\sum_{i}\frac{\partial V}{\partial\phi^i}\frac{d\phi^i}{d\varphi}\equiv\frac{dV}{d\varphi}, \\
 \label{eomproj2} &&\left(\frac{d^2\phi^i}{d\varphi^2}+\Gamma^i_{jk}\frac{d\phi^j}{d\varphi}\frac{d\phi^k}{d\varphi}\right)\dot{\varphi}^2=\left(g^{ik}-\frac{d\phi^i}{d\varphi}\frac{d\phi^k}{d\varphi}\right)\frac{dV}{d\phi^k}, 
\ea
also, \eqref{multi2} can be rewritten as
\be \label{einsteinproj} \Lambda=-\frac{1}{2}\dot{\varphi}^2+V, \ee
combining equations \eqref{eomproj1}, and \eqref{einsteinproj}, we get 
\be \label{multi4} \left(\frac{d\Lambda}{d\varphi}\right)^2=\alpha\Lambda\left(\Lambda-V\right), \ee
which is similar the equation we had for one direction. We again choose the negative sign for $\partial_{\varphi}\Lambda$
\be \frac{d\Lambda}{d\varphi}=-\sqrt{\alpha\Lambda\left(\Lambda-V\right)}. \label{multi3} \ee

The proposed distance generalizes to
\ba \Delta &=&\int \sqrt{g^{ij}(\phi^k)\left(1-\frac{V}{\Lambda}\right)d\phi_i d\phi_j},\\
&=&\int \sqrt{\left(1-\frac{V}{\Lambda}\right)}d\varphi, \ea
incorporating equation \eqref{multi3}, we get
\ba \Delta
&=&\frac{1}{\sqrt{\alpha}}\text{Log}\left(\frac{\Lambda_f}{\Lambda_i}\right). \ea
we therefore recover the AdS distance conjecture in multi-field case.

To define $\Lambda(\phi)$ at each point in the field space, we begin at a base point, $\phi^i_0$, with some initial $\Lambda(\phi^i_0)=\Lambda_0$. Subsequently, we can move to any arbitrary point in the field space and assign the value of $\Lambda$ at that point as $\Lambda(\phi^i)$.  Note that this definition of $\Lambda(\phi^i)$ thus defined will depend on the choice of both a base point and an initial value for $\Lambda$.

\section{ADM mass as a distance}\label{app:B}
In this Appendix, we develop the idea discussed in the main text of using the ADM mass of an asymptotically flat or AdS configuration as a notion of distance.
Specifically, consider the case $d=4$, and again the problem of defining a distance between two vacua with energies $V_1$ and $V_2$. To do this, we will look at the ADM mass of a generic bubble of radius $R$ (meaning it bounds a sphere of area $4\pi R^2$ interpolating between the two geometries). On general grounds, we expect that, for general $R$, the ADM mass will go as
\begin{equation} M\sim 4\pi\, TR^2-(\Delta V) \text{Vol}(R),\end{equation}
where $\text{Vol}(R)$ is the volume of a sphere of radius $R$. From this we can determine $T$, in the limit $R\rightarrow\infty$ this will define yet another notion of distance that will reduce to the domain wall tension. 

We will not solve the equations of motion, and look only at initial conditions at a moment of time-reversal symmetry. In that case, we only need to solve the Hamiltonian constraint from general relativity, which takes the form 
\begin{equation}G^{00}=8\pi G\, T^{00}.\end{equation}
Specifically, consider a transition from vacuum $V_{\text{in}}$ to $V_{\text{out}}$ for a scalar field $\phi$. We will take a metric ansatz
\begin{equation} ds_3^2= \left(1-\frac{8\pi GV(\phi)}{3}r^2-\frac{2Gm(r)}{r}\right)^{-1}dr^2+ r^2d\Omega_3^2\label{rr5}.\end{equation}
 The ADM mass of the configuration is simply $M(\infty)$ in the AdS/flat space cases (and presumably, $M$ evaluated at the horizon at the de Sitter case). With this metric ansatz, the Hamiltonian constraint takes the form (for spherically symmetric configurations)
\begin{equation}\frac{2 G \dot{m}}{r^2}+\frac{8}{3} \pi  G r V'\dot{\phi}+8 \pi  G V= 8\pi G\left(\frac12\dot{\phi}^2+V\right),\end{equation}
which simplifies to
\begin{equation}\frac{ \dot{m}}{4\pi\, r^2}+\frac{r}{3}  V'\dot{\phi}= \frac12\dot{\phi}^2.\end{equation}
Note that our ansatz allows us to get rid of the pure potential terms, similarly to what happens in the classical Maupertuis action. 

This is a first-order ODE for $\dot{m}$, which allows one to find an integral expression for the $T$, 
\begin{equation} TR^2=\int_R^\infty dr\,  \dot{m}(r)=4\pi \int_R^\infty dr\,r^2\left[\frac12\dot{\phi}^2-\frac{r}{3}V'\dot{\phi}\right] .\end{equation}
Now that $m(r)$ has been eliminated from the problem, our task is simply to find the on-shell value of the action $TR^2$ at an extremum, with the boundary conditions that $\phi=\phi_1$ at $r=R$, and $\phi=\phi_2$ at $R\rightarrow\infty$. We can integrate by parts
\begin{equation} \int_R^\infty dr\,r^3V'\dot{\phi}= Vr^3\vert^{\infty}_{R}-3\int r^2V,\end{equation}
to obtain
\begin{equation} TR^2+ \text{Vol}(R) \Delta V=\int_R^\infty dr\,  \dot{m}(r)=4\pi \int_R^\infty dr\,r^2\left[\frac12\dot{\phi}^2+V\right] - (Vr^3\vert^{\infty}_{R}) .\end{equation}
In other words, the final expression we obtain is again the classical expression for the energy  (while substracting a diverging boundary term if the asymptotic potential is nonzero). So we recover the non-gravitational problem. But not exactly; we need to check whether horizons form, i.e. whether we ever get that the $dr^2$ term in \eqref{rr5} changes sign. In other words, we need to impose
\begin{equation}m(r)\leq \frac{r}{2G}- \frac{4\pi}{3} Vr^3.\label{r334}\end{equation}
This is satisfied at $r=R$. Comparing derivatives, we get
\begin{equation} m'(r)=4\pi r^2\left[\frac12\dot{\phi}^2-\frac{r}{3}V'\dot{\phi}\right] \leq \frac{1}{2G} - \frac{4\pi }{3} V'\dot{\phi} r^3- 4\pi r^2\, V,\end{equation}
which rearranges to
\begin{equation} 4\pi r^2\left[\frac12\dot{\phi}^2+V\right] \leq \frac{1}{2G}. \end{equation}
If this satisfied, horizon never forms. If it is not, one may fix it by going to very large $R$, so that the rhs of \eqref{r334} is initially large, and so there is a lot of leeway to be satisfied. For fixed tension $T$, $TR^2$ is however larger than $R/G$ for  large enough $R$,  so unfortunately a horizon always forms in the limit we wish to take, unless perhaps in the AdS case. As described in the main text, this presents a hurdle for this notion being completely universal.

\bibliographystyle{jhep}
\bibliography{references}
\end{document}